%% file: anonymous-submission-latex-2025.tex
\documentclass[letterpaper]{article} 
\usepackage[draft]{aaai25}  
\usepackage{times}  
\usepackage{helvet}  
\usepackage{courier}  
\usepackage[hyphens]{url}  
\usepackage{graphicx} 
\usepackage{natbib}  
\usepackage{caption} 
\usepackage{algorithm}
\usepackage{algorithmic}

\usepackage{tikz}
\usetikzlibrary{positioning, shapes.geometric, arrows.meta}

\usepackage{newfloat}
\usepackage{listings}
\DeclareCaptionStyle{ruled}{labelfont=normalfont,labelsep=colon,strut=off} 
\floatstyle{ruled}
\newfloat{listing}{tb}{lst}{}
\floatname{listing}{Listing}
\input{preamble}

\title{DisMix: Disentangling Mixtures of Musical Instruments \\ for Source-level Pitch and Timbre Manipulation}

\title{DisMix: Disentangling Mixtures of Musical Instruments \\ for Source-level Pitch and Timbre Manipulation}
\author {
    Yin-Jyun Luo\textsuperscript{\rm 1}
    Kin Wai Cheuk\textsuperscript{\rm 2}
    Woosung Choi\textsuperscript{\rm 2}
    Toshimitsu Uesaka\textsuperscript{\rm 2}
    Keisuke Toyama\textsuperscript{\rm 3} \\
    Koichi Saito\textsuperscript{\rm 2}
    Chieh-Hsin Lai\textsuperscript{\rm 2}
    Yuhta Takida\textsuperscript{\rm 2}
    Wei-Hsiang Liao\textsuperscript{\rm 2}
    Simon Dixon\textsuperscript{\rm 1}
    Yuki Mitsufuji\textsuperscript{\rm 2,3}
}
\affiliations {
    \textsuperscript{\rm 1}C4DM, Queen Mary University of London \hspace{0,5em}
    \textsuperscript{\rm 2}Sony AI \hspace{0,5em}
    \textsuperscript{\rm 3}Sony Group Corporation\\
    yin-jyun.luo@qmul.ac.uk,
    kinwai.cheuk@sony.com
}

\begin{document}

\maketitle

\begin{abstract}
Existing work on pitch and timbre disentanglement has been mostly focused on single-instrument music audio, excluding the cases where multiple instruments are presented.
To fill the gap, we propose DisMix, a generative framework in which the pitch and timbre representations act as modular building blocks for constructing the melody and instrument of a source, and the collection of which forms a set of per-instrument latent representations underlying the observed mixture.
By manipulating the representations, our model samples mixtures with novel combinations of pitch and timbre of the constituent instruments.
We can jointly learn the disentangled pitch-timbre representations and a latent diffusion transformer that reconstructs the mixture conditioned on the set of source-level representations.
We evaluate the model using both a simple dataset of isolated chords and a realistic four-part chorales in the style of J.S. Bach, identify the key components for the success of disentanglement, and demonstrate the application of mixture transformation based on source-level attribute manipulation.

\end{abstract}

\input{sections/introduction}
\input{sections/related_work}
\input{sections/method}

\input{sections/vae}
\input{sections/vae-result}

\input{sections/ldm}

\input{sections/ldm-result}
\input{sections/conclusion}

%

\bibliography{bib}

\clearpage
\onecolumn

\appendix
\input{sections/appendix/appendix}

\end{document}

%% file: preamble.tex
\newcommand{\Dsymbol}[1] {\mathcal{D}_{\mathrm{#1}}}
\newcommand{\KLD}[2] {\Dsymbol{KL}\big(#1 \Vert #2\big)}
\newcommand{\gauss}[2]{\mathcal{N}\big(#1, #2\big)}
\newcommand{\Lo} {\mathcal{L}}

\newcommand{\E}[2] {\mathbb{E}_{#1}\big[ #2 \big]}

\graphicspath{ {figures/} }

\newcommand{\R}[1]{\mathbb{R}^{#1}}
\newcommand{\vcal}[1]{\mathcal{#1}}
\newcommand{\vrm}[1]{\mathrm{#1}}
\newcommand{\vi}[1]{#1^{(i)}}
\newcommand{\vj}[1]{#1^{(j)}}
\newcommand{\neqref}[1]{Eq.~\eqref{#1}}

\newcommand{\set}[2]{\{ #1^{(i)} \}_{i=1}^{#2}}

\newcommand{\Enc}[2] {\mathrm{E}_{\phi_{#1}}(#2)}
\newcommand{\qd}[2]{q_{\phi_#1}(#2)}
\newcommand{\pd}[2]{p_{\theta_#1}(#2)}

\usepackage{nccmath}
\usepackage{booktabs}
\usepackage{amsfonts}
\usepackage{mathtools}
\usepackage{xpatch}
\usepackage{icomma}
\usepackage{multirow}

%% file: sections/introduction.tex
\section{Introduction}\label{sec:introduction}

Disentangled representation learning (DRL) captures semantically meaningful latent features of observed data in a low-dimensional latent space~\cite{bengio_representation_2013}.
By applying a generative framework such as variational autoencoders (VAEs)~\cite{DBLP:journals/corr/KingmaW13}, we can train an encoder to encode the data and associate data-generating factors with separate subspaces in the latent space, and a decoder to reconstruct the data given the original encoding or to render novel data given manipulated latent features~\cite{tschannenRecentAdvancesAutoencoderBased,chenIsolatingSourcesDisentanglement2018,kimDisentanglingFactorising2018,higgins_beta-vae_2016}.

Because each feature representation lives in a subspace corresponding to a unique concept or data attribute such as the size or shape of a physical object in an image,
manipulating specific representations only renders variation of a few and particular factors in the decoder output, and thereby facilitates controllable transformation of existing data.

DRL has been applied to music audio to extract representations of timbre (e.g., the musical instrument played in a recording) and pitch (e.g., the melody played by the instrument)~\cite{luo_learning_2019,cifka_self-supervised_2021,tanaka_pitch-timbre_2021,luo_unsupervised_2020,tanaka_unsupervised_2022,luo_towards_2022,engelNeuralAudioSynthesis2017,bitton2018modulatedvariationalautoencodersmanytomany,wuTransplayerTimbreStyle2023,liuLearningInterpretableLowdimensional2023,wu2024emergentinterpretablesymbolscontentstyle}.
Disentangling the two attributes enables applications such as transferring the instrument played in a reference audio to a target instrument provided by another audio example, while preserving the melody played by the reference instrument.
This is similar to voice conversion, which aims at replacing a reference speaker's identity such that the converted speech sounds as if a different speaker spoke the content originally uttered by the reference speaker~\cite{hsuUnsupervisedLearningDisentangled2017,qianAutoVCZeroShotVoice2019}.


Despite being widely adopted for instrument attribute transfer, pitch-timbre disentanglement has been mostly applied to single-instrument audio.
Therefore, the analysis and the synthesis of the two attributes are only amenable to a single instrument at a time, which excludes cases where multiple instruments are presented in music audio.

Alternatively, one can use off-the-shelf source separation models to first separate each instrument from a mixture and apply the aforementioned approaches for disentanglement.
However, it introduces artefacts into the separated instrument and limits the instruments that can be handled to those supported by the state-of-the-art source separation models (e.g., drums, bass, vocals, and others)~\cite{lu2024sourceseparation,rouard2023TransformerDemucs,spleeter2020}.

To fill the gap, we propose $\textit{DisMix}$, a framework which \underline{dis}entangles a \underline{mix}ture of instruments and renders a novel mixture conditioned on a set of pitch-timbre disentangled representations.
DisMix represents each instrument, or source, in a mixture by a source-level representation that combines a pair of pitch and timbre latent variables.
The two attributes are encoded separately so that they can be independently manipulated for each instrument.
A decoder then takes as input the manipulated set of source-level representations and renders a new mixture that consists of sources whose pitch and timbre are dictated by the manipulated source-level representations.


The pitch and timbre representations act as modular building blocks to construct the melody and instrument of a source or an ``audio object''.
The decoder assembles these objects in a way that preserves their pitch and timbre.
This is reminiscent of ``object representation'' motivated by humans' capability to understand complex ideas in terms of reusable and primitive components~\cite{Greff2020OnTB,bizleyWhatWhereHow2013}.

To extract source-level representations, the pitch and timbre encoders are conditioned on a mixture and queries. The timbre of each query determines which instrument's latents are extracted; the extracted representation should match the timbre of the query, regardless of their pitch information. 

We propose two parameterisations of DisMix and evaluate using the MusicSlot~\cite{gha_unsupervised_2023} and the CocoChorale dataset~\cite{wu2022chamber}.
Our final implementation admits a latent diffusion model (LDM)~\cite{rombachHighResolutionImageSynthesis2022} which jointly learns the disentangled representations and a conditional diffusion transformer (DiT)~\cite{peeblesScalableDiffusionModels2023a}.
Our contributions are summarised as follows:
\begin{itemize}
    \item We propose DisMix to disentangle pitch and timbre of constituent sources from multi-instrument mixtures, whereby it is capable of rendering novel mixtures by manipulating the attributes of individual instruments.
    
    \item 
    Without employing domain-specific data augmentation or adversarial training, we propose and identify a binarisation layer as a key component for disentanglement, and apply the model to the four-part Bach Chorales from CocoChorale, which features 13 orchestral instruments.
    
    \item Based on the DiT architecture, we demonstrate that reconstructing a mixture conditioned on a set of jointly learnt source-level latents can yield superior performance than iteratively reconstructing a single source at a time.
    
\end{itemize}


%% file: sections/related_work.tex
\section{Related Work}\label{sec:related_work}

\subsubsection{Pitch and Timbre Disentanglement}

%

Strategies for pitch-timbre disentanglement include supervised learning~\cite{luo_learning_2019, engelNeuralAudioSynthesis2017,bitton2018modulatedvariationalautoencodersmanytomany,wuTransplayerTimbreStyle2023}, metric learning based on domain knowledge~\cite{esling2018generative,tanaka_pitch-timbre_2021, tanaka_unsupervised_2022,cifka_self-supervised_2021,luo_unsupervised_2020}, and more general inductive biases~\cite{cifka_self-supervised_2021,luo_towards_2022,liuLearningInterpretableLowdimensional2023,wu2024emergentinterpretablesymbolscontentstyle,luo2024jacobian}.
Despite their success for tasks such as attribute swapping, these methods are focused on single-instrument input.

Instead, we disentangle mixtures of instruments and represent each instrument by a source-level latent representation which captures pitch and timbre in separate dimensions.

Only a few studies explicitly extract pitch and timbre information from mixtures of instruments. \citet{hungMusicalCompositionStyle2019} and \citet{cwitkowitz2024timbre} encode the overall timbre of a mixture and consider applications of symbolic music rearrangement and transcription, respectively.
~\citet{cheuk2023jointistsimultaneousimprovementmultiinstrument} propose a multi-task framework that conditions music pitch transcription on intermediate timbre information.

~\citet{lin_unified_2021} tackle transcription and separation of mixtures and single-instrument generation in a unified framework.
Rather than focusing on single-instrument cases, we are interested in sampling novel mixtures by conditioning a DiT on a set of per-instrument representations and we evaluate our model by source-level attribute manipulation.

\subsubsection{Object-Centric Representation Learning}
Learning object representations entails encoding object entities in a visual scene by unique representations~\cite{Greff2020OnTB,locatelloObjectCentricLearningSlot2020}.
\citet{singhNeuralSystematicBinder2022,wuNeuralLanguageThought2023,wu2024neural} take a step further to disentangle attributes of individual objects for achieving compositional scene generation.
For audio, \citet{gha_unsupervised_2023} and \citet{audioslot2023AudioSlot} learn separate representations for different sources from mixtures.
We further disentangle pitch and timbre of individual sources and tackle a more complex dataset with a conditional LDM.






%% file: sections/method.tex
\begin{figure}
    \centering
    \includegraphics[width=0.85\linewidth]{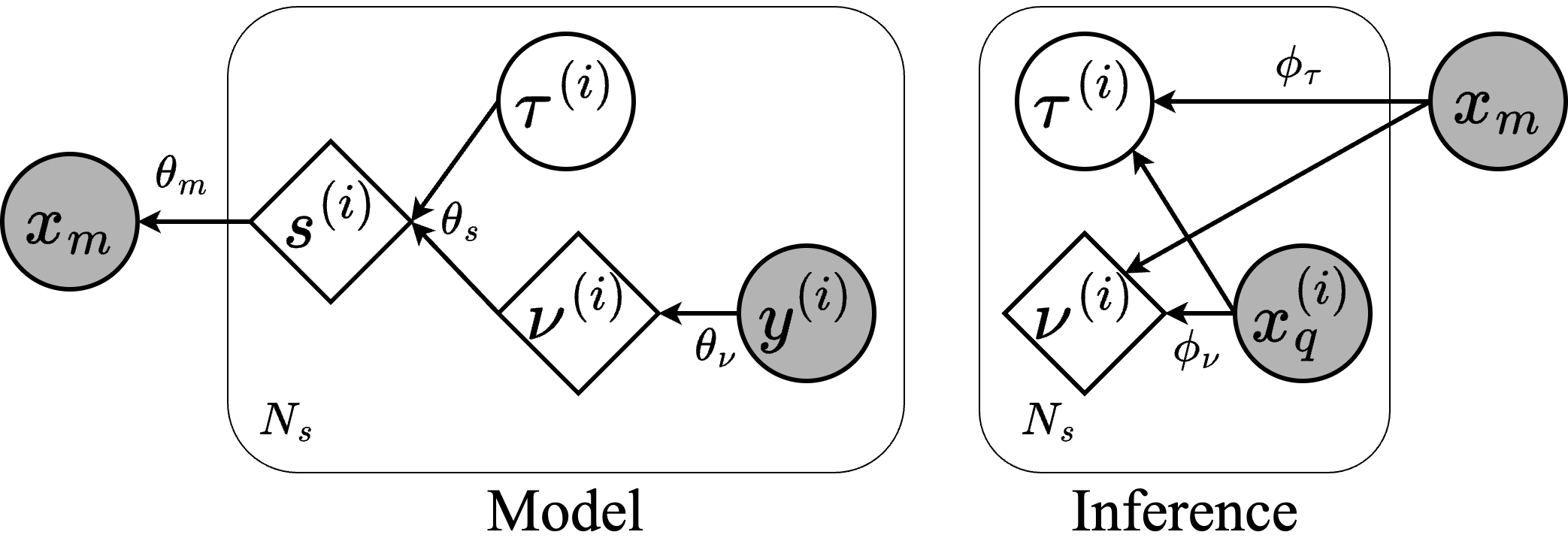}
    \caption{
    A mixture $x_m$ of $N_s$ instruments is represented by a set of source-level  latents $\set{s}{N_s}$ integrating latents of pitch $\vi{\nu}$ and timbre $\vi{\tau}$.
    Diamond nodes denote deterministic mappings.
}
    \label{fig:plate}
\end{figure}

\section{DisMix: The Proposed Framework}
Given a mixture $x_m$ of $N_s$ instruments and a query $\vi{x_q}$ (detailed in Section~\ref{sec:mixture_and_query}), our goal is to extract latent representations of pitch $\vi{\nu}$ and timbre $\vi{\tau}$ for an instrument $i \in [1, N_s]$, and to be able to sample a novel mixture conditioned on a set of manipulated pitch and timbre latents.

We propose DisMix to achieve the goal.
Its generative process on the left side of Fig.~\ref{fig:plate} samples $x_m$, through a neural network $\theta_m$, conditioned on a set of source-level representations $\vcal{S} = \set{s}{N_s}$ where $\vi{s} = \vrm{f}_{\theta_s}(\vi{\tau}, \vi{\nu})$ is a deterministic function of timbre and pitch.
The pitch latent $\vi{\nu}$ is computed via a neural network $\theta_\nu$ by its corresponding ground-truth pitch annotation $\vi{y}$.
The joint distribution over the variables can be written as follows:
\begin{equation}\label{eq:generative-process}
\pd{m}{x_m | \vcal{S}}
\prod\nolimits_{i=1}^{N_s}
\pd{\nu}{\vi{\nu}| \vi{y}}
p(\vi{\tau}),
\end{equation}
where $p(\vi{\tau}) = \gauss{0}{\vrm{I}}$.
Note that $\vi{\nu}$ and $\vi{\tau}$ are sampled independently and so are $\set{s}{N_s}$, whereby the pitch and timbre latents can act as modular building blocks to compose a mixture.

We propose two instances of DisMix: 1) an auto-encoder for a simple case study using a simplistic dataset (Section~\ref{sec:instance_1}) and 2) an LDM implemented using a DiT verified with a more complex dataset (Section~\ref{sec:instance_2}).
We specify $\theta_m$, $\theta_s$, and $\theta_\nu$ for the two instances in Sections~\ref{sec:instance_1} and~\ref{sec:instance_2}, respectively.
Here, we describe the common components between them.



Following VAEs~\cite{DBLP:journals/corr/KingmaW13}, we learn DisMix using an inference network to approximate posteriors over the latent variables that are otherwise intractable, 
and learns the model by maximising a lower bound to the marginal likelihood $p(x_m | \set{y}{N_s})$.
For the two instances of DisMix, we employ a common parameterisation of the inference network illustrated on the right side of Fig.~\ref{fig:plate}, which we explain in the following sections.

\subsection{Mixture and Query Encoders}\label{sec:mixture_and_query}


Given a mixture of instruments, additional information would be necessary to specify which instrument's latents are to be extracted. 
Motivated by~\citet{lin_unified_2021} and~\citet{lee_audio_2019}, we use the query $\vi{x_q}$ so that the extracted latents of the source $i$ and the query share the same timbre characteristics, while they can carry arbitrary pitch information.
We describe the objective function~\neqref{eq:bt} that could encourage this behaviour in Section~\ref{sec:training_object}.

During training, we pair each mixture $x_m$ with $\vcal{X}_q = \set{x_q}{N_s}$, a set of $N_s$ queries with their instruments corresponding to the constituent instruments of the mixture, and each $\vi{x_q}$ is randomly sampled from a subset of the complete training dataset consisting of a constituent instrument.

As illustrated on the right side of Fig.~\ref{fig:plate}, the inference of $\vi{\tau}$ and $\vi{\nu}$ is conditioned on both $x_m$ and $\vi{x_q}$.
In practice, we extract their compact counterparts $e_m = \Enc{m}{x_m}$ and $\vi{e_q} = \Enc{q}{\vi{x_q}}$, respectively, where $\Enc{m}{\cdot}$ and $\Enc{q}{\cdot}$ are neural network encoders with parameters $\phi_m$ and $\phi_q$.

\subsection{Pitch and Timbre Encoders}\label{sec:disentanglement}
As suggested by Fig.~\ref{fig:plate}, we propose that the pitch and timbre encoders admit a common factorised form:
\begin{equation}\label{eq:posterior}
\qd{u}{\vcal{U} | x_m, \vcal{X}_q} = \prod\nolimits_{i=1}^{N_s} \qd{u}{\vi{u} | e_m, \vi{e_q}},
\end{equation}
where $u \in \{\nu, \tau\}$ and $\vcal{U} \in \{\set{\nu}{N_s}, \set{\tau}{N_s}\}$. 
$\phi_\nu$ and $\phi_\tau$ are parameters of the two encoders.
Given $e_m$ and $\vi{e_q}$, the pitch and timbre latents of the $i$-th source are encoded independently of each other and of other sources.

We apply a binarisation layer~\cite{dong_convolutional_2018} to constrain the capacity of the pitch latent, which is proven crucial for disentanglement in our empirical studies.
We also show that imposing the standard Gaussian prior $p(\vi{\tau})$ is a simple yet effective way to constrain the timbre latent.


\subsubsection{Constraining Pitch Latents}
The pitch encoder combines a transcriber $\Enc{\nu}{\cdot}$ that extracts pitch logits of the $i$-th source $\vi{\hat{y}} = \Enc{\nu}{e_m, \vi{e_q}}$, a stochastic binarisation layer ($\vrm{SB}$) that constrains the information capacity, and a translator $\vrm{f}_{\theta_\nu}(\cdot)$ that computes the final outcome:
\begin{equation}\label{eq:pitch-encoder}
\qd{\nu}{\vi{\nu} | e_m, \vi{e_q}} \coloneqq \delta(\vi{\nu} - \mathrm{f}_{\phi_\nu}(\vi{\hat{y}_{\vrm{bin}}})),
\end{equation}
where $\delta(\cdot)$ is the Dirac delta function, corresponding to the diamond node of $\vi{\nu}$ on the right of Fig.~\ref{fig:plate}, and:
\begin{equation}\label{eq:y_bin}
   \vi{\hat{y}_{\vrm{bin}}} = \vrm{SB}(\vi{\hat{y}}) = \mathbf{1}_{\{\vrm{Sigmoid}(\vi{\hat{y}}) > h\}},
\end{equation}
where $\mathbf{1}_{\{\cdot\}}$ is the indicator function, and
$h$ is the threshold sampled from the uniform distribution $\vrm{U}(0,1)$ at each training step and is fixed at $0.5$ during evaluation.
The straight-through estimator~\cite{Bengio2013EstimatingOP} is used to bypass the non-differentiable operator.


Table~\ref{tab: loss-ablation} suggests that the bottleneck imposed by the binarisation layer is crucial for disentanglement even if a pitch classification loss is included.
We also show that using only $\vrm{SB}$ without the pitch supervision still yields a decent performance in Table~\ref{tab:ldm_result}.

\subsubsection{Constraining Timbre Latents}
The timbre encoder parameterises a Gaussian with a neural network $\phi_\tau$:
\begin{equation}
\qd{\tau}{\vi{\tau}} = \mathcal{N}(\vi{\tau} ;\mu_{\phi_{\tau}}(e_m, \vi{e_q}), \sigma_{\phi_{\tau}}^2(e_m, \vi{e_q})\vrm{I}),
\end{equation}
where $\qd{\tau}{\vi{\tau}} \coloneqq \qd{\tau}{\vi{\tau} | e_m, \vi{e_q}}$ and sampling $\vi{\tau}$ is reparameterised as $\mu_{\phi_{\tau}}(\cdot) + \epsilon \sigma_{\phi_{\tau}}(\cdot)$, where $\epsilon \sim \gauss{0}{\vrm{I}}$, to be differentiable~\cite{DBLP:journals/corr/KingmaW13}. 

We let $p(\vi{\tau}) = \gauss{0}{\vrm{I}}$ to constrain the timbre latent
through the Kullback–Leibler divergence (KLD) in~\neqref{eq:elbo2}.

\subsection{Training Objectives}\label{sec:training_object}
\subsubsection{ELBO} We start with an evidence lower bound (ELBO) to the marginal log-likelihood $\log p(x_m | \set{y}{N_s})$:
\begin{equation}\label{eq:elbo2}
\begin{split}
   \vcal{L}_{\vrm{ELBO}} = 
   \E{
 \prod\nolimits_i \qd{\tau}{\vi{\tau}}
 }{
 \log \pd{m}{x_m | \{\vi{\tau}, \vi{\hat{\nu}} \}_{i=1}^{N_s} }
}  \\ +
\sum\nolimits_i
\log\pd{\nu}{\vi{\hat{\nu}} | \vi{y}} -
\KLD{
\qd{\tau}{\vi{\tau}}
}{
p(\vi{\tau})
},
\end{split}
\end{equation}
where $\vi{\hat{\nu}} = \mathrm{f}_{\phi_\nu}(\vi{\hat{y}_{\vrm{bin}}}$) by~\neqref{eq:pitch-encoder}.
Note that the conditionals of the first term can be expressed in terms of $\vcal{S} = \set{s}{N_s}$, where $\vi{s} = \vrm{f}_{\theta_s}(\vi{\tau}, \vi{\hat{\nu}})$ as described previously for \neqref{eq:generative-process}.
$\vcal{L}_{\vrm{ELBO}}$ is derived as a result of the factorised posteriors in~\neqref{eq:posterior} and the deterministic pitch encoder by~\neqref{eq:pitch-encoder}, which we detail in Appendix~\ref{app:elbo}.
We specify $\theta_m$, $\theta_s$, and $\theta_\nu$ in Section~\ref{sec:instance_1} and~\ref{sec:instance_2}.

Intuitively speaking, we extract both a pitch and a timbre latent for each source given a mixture and a query, and the collection of which is used to reconstruct the mixture.
The pitch and timbre latents adhere to certain constraints and priors to encourage the disentanglement of the two attributes.

Apart from mixtures, we also observe individual sources $\vi{x}_s$, so we include a source-wise reconstruction loss.
This is a special case of the first term in~\neqref{eq:elbo2} when $N_s =1$ and $x_m$ becomes $\vi{x_s}$, and we can reuse $\theta_m$ to reconstruct individual sources.

\subsubsection{Pitch Supervision}\label{sec:auxliary_loss}
To enhance the disentanglement, we also minimise a binary cross entropy loss $\vrm{BCE}(\vi{\hat{y}}, \vi{y})$ where $\vi{y}$ is the pitch annotation of the $i$-th source.
We explore relaxing the model by excluding this term, with the results reported in Table~\ref{tab:ldm_result}.

\subsubsection{Barlow Twins}
Finally, we minimise a simplified Barlow Twins loss~\cite{zbontar_barlow_2021} to enhance the correlation between the query and the timbre latent for them sharing the timbre characteristics as described in Section~\ref{sec:mixture_and_query}:
\begin{equation}\label{eq:bt}
\Lo_{\mathrm{BT}} = \sum\nolimits_{i=1}^{N_s}\sum\nolimits_{d=1}^{D_\tau} (1 - \mathcal{C}_{dd}(\vi{e}_q, \vi{\tau}))^2,
\end{equation}
where $\mathcal{C}$ is a cross-correlation matrix, and both $\vi{e}_q$ and $\vi{\tau}$ share the same dimensionality $D_\tau$.
Empirically, $\vcal{L}_{\vrm{BT}}$ counteracts the over-regularisation effect of the prior $p(\vi{\tau})$ and promotes a discriminative timbre space as shown in Fig.~\ref{fig:latent-and-spec}.

\subsubsection{The Final Objective}
In summary, we maximise:
\begin{equation}\label{eq:dismix-loss}
\Lo_{\vrm{DisMix}} = \Lo_{\vrm{ELBO}}  -\Lo_{\vrm{BCE}} - \Lo_{\vrm{BT}}.
\end{equation}
We do not find explicitly weighting each loss term necessary.

Next, we detail the implementations specific to the simple and more sophisticated variants of DisMix in Sections~\ref{sec:instance_1} and~\ref{sec:instance_2}, respectively.

%% file: sections/vae.tex
\section{A Simple Case Study}\label{sec:instance_1}

\subsubsection{Reconstructing Mixtures}
$\pd{m}{x_m | \vcal{S}}$ in ~\neqref{eq:elbo2} is a Gaussian likelihood parameterised by a decoder $\theta_m$, which can be a permutation invariant function such as a transformer~\cite{vaswani_attention_2017} without positional embeddings that outputs the Gaussian mean, reconstructing a mixture given a set of source-level representations.

We opt for a simple implementation of $\theta_m$ here and discuss a transformer in Section~\ref{sec:instance_2}.
In particular, we slightly deviate from~\neqref{eq:elbo2} and reconstruct $x_m$ using $e_m$ in Section~\ref{sec:mixture_and_query}, instead of $\vcal{S}$, whereby the likelihood becomes $\pd{m}{x_m | e_m}$, and we maximise an additional term:
\begin{equation}\label{eq:mixture_prior}
\begin{split}
\E{\prod\nolimits_i \qd{\tau}{\vi{\tau}}}{\log p(e_m | \{\vi{\tau}, \vi{\hat{\nu}} \}_{i=1}^{N_s}},
\end{split}
\end{equation}
where
$p(e_m | \vcal{S}) = \mathcal{N}(e_m; \sum\nolimits_{i=1}^{N_s}\vi{s}, \sigma_m^2 \vrm{I} )$ and $\sigma_m = 0.25$ is a hyperparameter.
Intuitively, we extract the source-level latents $\vcal{S} = \set{s}{N_s}$ whose summation $s_{\vrm{sum}} = \sum_i \vi{s}$ and $e_m$ are pulled together as measured by a mean square error weighed by $\sigma_m^{-2}$.

During evaluation, we instead use $s_\vrm{sum}$ to reconstruct the mixture $x_m$ or render a novel one by manipulating $\set{s}{N_s}$ before the summation.
The assumption is that both $e_m$ and $s_{\vrm{sum}}$ can reconstruct $x_m$ comparably well by maximising~\neqref{eq:mixture_prior} as we reconstruct $x_m$ by $e_m$ during training.

This approach avoids implementing a (potentially expensive and complicated) permutation invariant decoder
and imposes linearity between $e_m$ and $s_{\vrm{sum}}$ in the latent space.
The linearity could enable other applications, which we explain in Appendix~\ref{app:ae_lin} and leave for future work.

\subsubsection{Integrating Pitch and Timbre}
We employ $\vrm{FiLM}$~\cite{perez_film_2018,kim_neural_2018} and derive the source-level representation as $\vi{s} = \vrm{f}_{\theta_m}(\vi{\tau}, \vi{\nu}) = \alpha_{\theta_s}(\vi{\tau}) \odot \vi{\nu} + \beta_{\theta_s}(\vi{\tau})$, 
where $\vi{\nu}$ is scaled and shifted element-wise by the factors determined as a deterministic function $\theta_s$ of $\vi{\tau}$.

\subsubsection{Configuring Pitch Priors}\label{sec:pitch-prior}
We study pitch priors at different levels of capacity.
The first follows~\neqref{eq:generative-process} and defines
$\pd{\nu}{\vi{\nu} | \vi{y}} = 
\mathcal{N}(\vi{\nu}; \mu_{\theta_\nu}^{\vrm{fac}}(\vi{y}), \sigma_{\nu}^2 \vrm{I})$, a Gaussian parameterised by $\theta_\nu$ given the ground-truth pitch $\vi{y}$.
$\sigma_\nu$ is a hyperparameter.

We also consider a richer prior to capture the source interaction:
$\pd{\nu}{\vi{\nu} | \vcal{Y}_{\backslash{i}}} =  \sum\nolimits_{k=1}^K 
 \pi_k \gauss{\vi{\nu} ;\mu_{\theta_\nu, k}^{\vrm{rich}}(\vcal{Y}_{\backslash{i}})}{ \sigma_{\nu}^2\vrm{I}}$,
where $\vcal{Y}_{\backslash{i}}$ denotes a set of pitch annotations excluding that of the $i$-th source, and $\theta_{\nu, k}$ parameterises the mean of the $k$-th Gaussian in a Gaussian mixture.
The rationale is that $\vi{\hat{y}}$ is conditionally dependent on pitch of other sources in a mixture to confer musical harmony.

\subsection{Implementations}\label{sec:musicslot}
\subsubsection{Dataset}
\citet{gha_unsupervised_2023} compile synthetic audio using 3,131 unique chords from JSB Chorales~\cite{boulanger-lewandowski_modeling_nodate}, rendered by sound fonts of piano, violin, and flute via FluidSynth.

Given a chord, each composite note is synthesised to an audio waveform at 16kHz with a sound font randomly sampled with replacement, whereby a sound font $i$ can play multiple notes in a chord. 
These notes together define a source $\vi{x_s}$ whose pitch annotation is a mutli-hot vector $\vi{y} \in \{0, 1\}^{N_p}$.
The note waveforms are summed to form the chord's waveform which defines a mixture $x_m$.

There are $28,179$ samples of mixtures split into the train, validation, and test sets with a ratio of $70$/$20$/$10$.
The waveforms are converted into mel spectrograms using $128$ mel-filter bands, a window size of $1,024$, and a hop length of $512$.
We crop a $320$ms segment, or $10$ spectral frames, from the sustain phase of each sample. 

\subsubsection{Architecture}
Simple layers including the $\vrm{MLP}$ and $\vrm{RNN}$ are used for implementation to evaluate this simple case study, and we extend DisMix to a more complex setup with an LDM in Section~\ref{sec:instance_2}.
A major difference between the two setups is that both $\vi{\nu}$ and $\vi{\tau}$ are represented as a single vector, as pitch and timbre are time-invariant within the simplified scope of the MusicSlot dataset~\cite{gha_unsupervised_2023}.
On the other hand, the CocoChorale dataset~\cite{wu2022chamber} features time-varying melodies in each sample.
We elaborate the implementation details in Appendix~\ref{app:ae}.

%
%
%

\subsubsection{Optimisation}
We use Adam~\cite{kingma_adam_2017} and a batch size of $32$, a learning rate of $0.0004$, and a gradient clipping value of $0.5$.
Training is terminated if~\neqref{eq:dismix-loss} stops improving on the validation set for $260$k steps.

%% file: sections/vae-result.tex

\subsection{Results}\label{sec:result_1}
\subsubsection{Evaluation}
Given $x_m$ and a set of queries $\set{x_q}{N_s}$ corresponding to the constituent instruments of the mixture, we first extract $\{\vi{\tau}, \vi{\nu}\}_{i=1}^{N_s}$.
To evaluate disentanglement, we conduct a random permutation so that the pitch latent of the source $i$ can be swapped for that of the source $j$, while the timbre remains unchanged, which yields $\vi{\hat{s}} = \vrm{f}_{\theta_s}(\vi{\tau}, \vj{\nu})$.
Then, we render a novel source $\vi{\hat{x_s}}$ by passing $\vi{\hat{s}}$ to the decoder $p_{\theta_m}$.
Note that we render sources instead of mixtures by using $\vi{\hat{s}}$ as the input instead of the summation of multiple source-level representations.

A successful disentanglement entails that judges of pitch and instrument should classify $\vi{\hat{x}_s}$ as the pitch of $\vj{x_s}$ (as it was swapped) and the instrument of $\vi{x_s}$ (as it was preserved), respectively.
We pre-train a pitch and an instrument classifier using the training set and direct them as the judges.
The classification accuracy is reported under ``Disentanglement'' in Table~\ref{tab: loss-ablation} and~\ref{tab: pitch-param}.

To see if the model can render novel mixtures, we first produce a new set $\set{\hat{s}}{N_s}$ after the permutation and render a novel mixture $\hat{x}_m$ by passing $p_{\theta_m}$ the summation $\hat{s}_{\vrm{sum}} = \sum_i \vi{\hat{s}}$.
To check whether the constituent attributes of $\hat{x}_m$ are indeed dictated by the manipulated $\set{\hat{s}}{N_s}$, we once again extract its source-level representations with the original queries and reconstruct the sources, which are then fed to the judges.
``Mixture Rendering'' in Table~\ref{tab: loss-ablation} and~\ref{tab: pitch-param} report the classification accuracy.

\begin{table}[t]
\small
\centering
\begin{tabular}{lrrrr}
\toprule
   & \multicolumn{2}{c}{Disentanglement} & \multicolumn{2}{c}{Mixture Rendering} \\ \cmidrule(lr){2-3} \cmidrule(lr){4-5}
   & Pitch             & Inst.           & Pitch               & Inst.            \\ \midrule
DisMix &  93.39\%           & 100.00\%           & 90.69\%             & 100.00\%            \\
- $\Lo_{\mathrm{BT}}$ & 93.18\%           & 99.92\%           & 87.92\%             & 100.00\%            \\ 
- KLD & 69.41\%           & 100.00\%           & 35.10\%             & 100.00\%            \\
- $\mathrm{SB}$ & 93.46\%           & 46.71\%           & 40.23\%             & 98.91\%            \\


\bottomrule
\end{tabular}
\caption{Ablation study: classification accuracy for various loss terms for the simple case study. 
}
\label{tab: loss-ablation}
\end{table}

\begin{figure}[ht]
    \centering
    \includegraphics[width=0.9\linewidth]{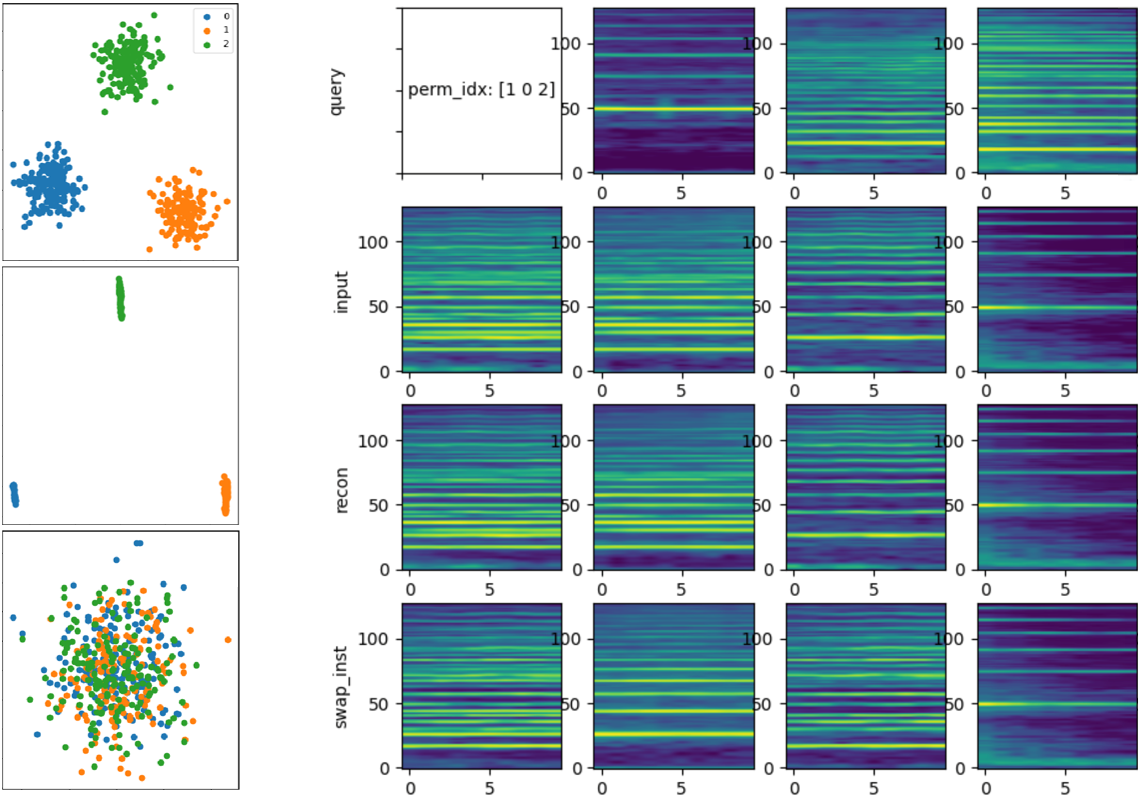}
    \caption{
    \textit{Left}: PCA of the timbre space. 
    Top: DisMix, plot $\vi{\tau}$. Mid and bottom: Remove $\Lo_{\mathrm{BT}}$, plot the mean of $\qd{\tau}{\vi{\tau}}$ 
    and the sampling, respectively.
    \textit{Right}: Novel mixture rendering.
    Refer to Section~\ref{sec:mixture-gen} for details.}
    \label{fig:latent-and-spec}
\end{figure}

\subsubsection{Examples}\label{sec:mixture-gen}
Fig.~\ref{fig:latent-and-spec} shows mel spectrograms on the right side.
The first row shows queries for different instruments.
The second row shows an input mixture $x_m$ and the corresponding sources.
The timbre characteristics are consistent across rows for the last three columns.
In the third row, the leftmost is the reconstructed mixture given $s_{\vrm{sum}}$ and the rest are the reconstructed sources given $\vi{s}$.

The first column of the last row refers to the manipulated mixture rendered by $\hat{s}_{\vrm{sum}}$, and the rest are the sources extracted from the mixture using the original queries.
Attributes are successfully manipulated. In the last row, the second column combines the first source's timbre with the second's pitch, and the third column combines the second's timbre with the first's pitch. The third source is unchanged.

\subsubsection{Ablation Study}
Table~\ref{tab: loss-ablation} is an ablation study of the loss term for identifying the key loss terms for disentanglement.
The table suggests that removing the KLD in~\neqref{eq:elbo2}, or the prior $p(\vi{\tau})$ from the timbre latent space, contaminates the timbre latent $\vi{\tau}$ with the pitch information.
Consequently, the pitch of $\vi{\hat{x}_s}$ is determined not only by $\vj{\nu}$, but also by $\vi{\tau}$, leading to a low pitch accuracy of $69.41\%$.

On the other hand, removing the $\vrm{SB}$ layer introduces excessive capacity for the pitch latent, such that the instrument of $\vi{\hat{x}_s}$ is dictated by both $\vi{\tau}$ and $\vj{\nu}$, resulting in a low instrument classification accuracy of $46.71\%$.
Even though the $\vrm{BCE}$ loss for pitch supervision is always included in the training, the model fails to disentangle without $\vrm{SB}$, suggesting that an additional information constraint is necessary in addition to supervision to achieve disentanglement.

In terms of mixture rendering, removing $\vcal{L}_{\vrm{BT}}$ negatively affects pitch accuracy.
This could be attributed to the timbre prior $p(\vi{\tau})$ which potentially over-regularises $\vi{\tau}$, leading to noisy samples of $\vi{\tau}$ due to a large posterior variance $\sigma_{\phi_\tau}^2(\cdot)$.
This could cause a poor optimisation of~\neqref{eq:mixture_prior}, aggravating the existing gap between using $e_m$ and $s_{\vrm{sum}}$ for mixture reconstruction during training and evaluation, respectively.
The left side of Fig.~\ref{fig:latent-and-spec} shows that, when $\Lo_{\mathrm{BT}}$ is removed, the instrument clusters in the timbre space disappear from the PCA of $\vi{\tau} \sim \qd{\tau}{\vi{\tau} | e_m, \vi{e_q}}$, suggesting excessively noisy sampling.

\begin{table}[!t]
\small
\centering
\begin{tabular}{l@{\hskip 5pt}c@{\hskip 5pt}c@{\hskip 5pt}c}
\toprule
   & DisMix & +fac & +rich ($K$=10) \\ 
\midrule
Disentanglement & 93.39\%  & 93.98\%  & 94.18\% \\
Mixture Rendering & 90.69\%  & 91.15\%  & 92.04\% \\
\bottomrule
\end{tabular}
\caption{Pitch accuracy using different pitch priors.}
\label{tab: pitch-param}
\end{table}

\subsubsection{Pitch Priors}

In Table~\ref{tab: pitch-param}, we compare the pitch priors discussed in Section~\ref{sec:instance_1} in terms of pitch accuracy.
It includes the model without a pitch prior ($\vrm{DisMix}$), with the factorised prior ($\vrm{fac}$), and with the one that captures the source interaction by a Gaussian mixture of $K=10$ components.
The standard deviation $\sigma_\nu$ is $0.5$ in both cases.
We can observe that the accuracy improves with a richer prior.

%% file: sections/ldm.tex
\section{A Latent Diffusion Framework}\label{sec:instance_2}
We also implement DisMix by an LDM framework~\cite{rombachHighResolutionImageSynthesis2022}, where the decoder $\pd{m}{x_m | \vcal{S}}$ is a diffusion transformer (DiT)~\cite{peeblesScalableDiffusionModels2023a} that directly reconstructs a mixture given a set of source-level latents, and no additional objective such as~\neqref{eq:mixture_prior} is introduced.

\subsubsection{Data Representation}
The LDM framework~\cite{rombachHighResolutionImageSynthesis2022} improves the compute efficiency of diffusion models (DMs)~\cite{sohl2015deep,ho2020denoising} by first projecting data to a low-dimensional latent space.
We leverage the pre-trained VAE from AudioLDM2~\cite{audioldm2-2024taslp} which is trained on multiple music and audio datasets to extract $\vi{z_s} = \vrm{E}_{\vrm{vae}}(\vi{x_s})$, where $\vrm{E}_{\vrm{vae}}(\cdot)$ is the VAE encoder, and we use the pre-trained VAE decoder $\vrm{D}_{\vrm{vae}}(\cdot)$ to recover $\vi{x_s}$.

\subsubsection{Latent Diffusion Models}
LDMs operate DMs in a latent space 
and sample $z_0$, the latent representation of data, from a Markov chain: $p(z_T)\prod\nolimits_{t=1}^{T} p_{\theta}(z_{t-1} | z_{{t}})$, with $p(z_T) = \vcal{N}(z_T;0,\vrm{I})$ and $p_{\theta}(z_{t-1}| z_t) = \mathcal{N}(z_{t-1};\mu_{\theta}(z_t, t), \Sigma_{\theta}(z_t, t))$ parameterised by $\theta$.
The posterior 
is a linear Gaussian $q(z_t | z_{t-1}) = \vcal{N}(z_t;\sqrt{\alpha_t}z_{t-1},(1-\alpha_t)\vrm{I})$, where $\alpha_t$ is a hyperparameter evolving over the diffusing step $t$.


Given that the forward process $q$ is known and fixed, \citet{ho2020denoising} employ specific forms of $\mu_{\theta}$ and $\Sigma_{\theta}$ to match $q$ and simplify the training to essentially minimising $\| \vrm{f}_{\theta}(z_t, t) - z_0\|^2_2$, which boils down to training a decoder $\vrm{f}_{\theta}$ to predict the clean $z_0$ given its corrupted counterpart $z_t$. 

\subsection{Adapting Diffusion Transformers}\label{sec:ldm_adapt}
Different from the vanilla LDMs, we condition the decoder with $\vcal{S}$, replacing the first term of~\neqref{eq:elbo2} with:
\begin{equation}\label{eq:ldm_decoder}
\pd{m}{z_{m, 0:T} | \vcal{S}} = 
p(z_{m, T})\prod\nolimits_{t=1}^T \pd{m}{z_{m, t-1} | z_{m, t}, \vcal{S}}.
\end{equation}
$z_{m, t}$ denotes the noised latent feature of mixtures at diffusing step $t$ and is defined in terms of $\set{z_{s, t}}{N_s}$, where $\vi{z_{s, 0}} = \vi{z_s}$.
$p_{\theta_m}$ models the interaction among the elements in the set $\vcal{S} = \set{s}{N_s}$ and iteratively performs the reverse process.
Defining $z_{m, t}$ in terms of the constituent sources facilitates the conditioning mechanism that we explain next.


\subsubsection{Partition}
DiTs~\cite{peeblesScalableDiffusionModels2023a} operate on a sequence of image patches to work with a transformer.
Similarly, we first apply a sinusoidal positional encoding to $\vi{z_s}$ to preserve the temporal order and partition it by
$\vrm{Par}(\vi{z_s}): \R{T_z \times (D_z \times C)} \rightarrow \R{L \times D_{z}'}$,
where $T_z=100$, $D_z=16$, $C=8$, and $L=25$ are the numbers of time frames, feature dimensions, channels, and patches, respectively.
That is, we partition along the time axis and flatten each patch whose size becomes $D_{z}' = \frac{T_z}{L} \times D_z \times C$.
We repeat the process for all $N_s$ elements in $\set{z_s}{N_s}$, the outcome of which finally defines $z_{m, t} \in \R{(N_s \times L) \times D_z'}$ in \neqref{eq:ldm_decoder}.
In other words, we can consider it as a sequence of $N_s \times L$ patches, with the size of each patch being $D_z'$.

Similarly, we partition the source-level representations $\vcal{S} = \set{s}{N_s}$ and obtain $ s_c \in \R{(N_s \times L) \times D_s'}$, where $D_s' = \frac{T_z}{L} \times D_s \times C$, which we detail in Section~\ref{sec:ldm_implementation}.
We can then represent the set condition $\vcal{S}$ by $s_c$ in~\neqref{eq:ldm_decoder}. 

As mentioned earlier, we define $z_{m,t}$ in terms of its constituent sources and conveniently align the dimensions of $s_c$ and $z_{m, t}$ (except for their sizes $D_z'$ and $D_s'$) which facilitates the conditioning described next.

We do not add another positional encoding to $z_{m, t}$ and $s_c$, ensuring the
permutation invariance w.r.t. the $N_s$ sources.

\subsubsection{Conditioning}
A transformer block consists of the multi-head self-attention mechanism and a feedforward network.
Each of these modules is followed by a skip connection and a layer normalisation~(LN) \cite{vaswani_attention_2017}.
We replace the standard LN with its adaptive variant (adaLN) for the conditioning~\cite{peeblesScalableDiffusionModels2023a}.

In particular, $s_c$ is added with a diffusing step embedding and is used to regress the scaling and shifting factors $\rho$ and $\gamma$ of adaLN layers which match the dimension of $z_{m, t}$.
Then, by passing $z_{m,t}$ through the transformer blocks, it is modulated by $\rho$ and $\gamma$ which carry the source-level pitch and timbre information computed from $s_c$.

\subsection{Implementations}\label{sec:coco}
\subsubsection{Dataset}
The CococChorale dataset~\cite{wu2022chamber} provides realistic generative music in the style of Bach Chorales~\cite{boulanger-lewandowski_modeling_nodate}, featuring $13$ 
orchestral instruments played at the pitch ranges of a standard four-part chorale (i.e., Soprano, Alto, Tensor, Bass, or SATB).

We use the random ensemble subset of $60,000$ four-part chorales, totalling $350$ hours of audio.
Each part of an example is played by an instrument randomly sampled from a pool of instruments belonging to the part.
The maximum number of sources $N_s$ is four.
The pitch annotation $\vi{y} \in \{0,1\}^{N_p \times T_p}$ is a time sequence of one-hot vectors.
The dataset follows a split ratio of $80$/$10$/$10$ for the training, validation, and test sets.

Each audio file is sampled at $16$kHz and is converted into mel spectrograms using $64$ mel-filter bands, a window size of $1,024$, and a hop length of $160$.
We randomly sample a four-second segment for each example during training.

\subsubsection{Architecture}\label{sec:ldm_implementation}

We reuse the architecture of the pre-trained VAE encoder $\vrm{E}_{\vrm{vae}}$ and decoder $\vrm{D}_{\vrm{vae}}$ from AudioLDM2~\cite{audioldm2-2024taslp} to implement parts of our encoders, which leads to the a pitch latent $\vi{\nu} \in \R{C \times T_z \times D_z}$ and a timbre latent $\vi{\tau} \in \R{C \times 1 \times D_z}$.
Note that we compute a time-invariant timbre latent to reflect the fact that the instrument identity does not change within each individual source.
The source-level representation $\vi{s} \in \R{C \times T_z \times D_s}$ simply concatenates $\vi{\nu}$ and $\vi{\tau}$, where $D_s = 2D_z$.
We can then apply the partition method to $\set{s}{N_s}$ and obtain $ s_c \in \R{(N_s \times L) \times D_s'}$.

By~\neqref{eq:ldm_decoder}, we can reconstruct $z_m$, defined in terms of $\set{z_s}{N_s}$, by sampling the DiT $p_{\theta_m}$ conditioned on the set $\vcal{S}$, represented as $s_c$.
We use HiFi-GAN~\cite{kong2020hifi} to convert the reconstructed mel spectrograms $\vrm{D}_{\vrm{vae}}(\vi{\hat{z}_s})$ back to audio $\vi{\hat{x_s}}$, and we can obtain the mixture by $\hat{x}_m = \sum\nolimits_{i=1}^{N_s}\vi{\hat{x}_s}$.
More implementation details are left in Appendix~\ref{app:ldm}

%
%

\subsubsection{Optimisation}
We use Adam~\cite{kingma_adam_2017} with a batch size of eight and a gradient clipping value of $0.5$.
The learning rate warms up linearly to $0.0001$ over $308$k steps and decreases following the cosine wave~\cite{loshchilov2017sgdr} for a maximum of $4,092$k steps.

%% file: sections/ldm-result.tex
\subsection{Results}
\subsubsection{Evaluation}
Given a reference mixture and its extracted set of pitch and timbre latents $\{\vi{\tau}, \vi{\nu}\}_{i=1}^{N_s}$, we arbitrarily replace the timbre latents $\set{\tau}{N_s}$ with the set from another target mixture.
We expect that the four instruments of the reference mixture are replaced by those of the target mixture and that the target instruments play the original melodies of the reference instruments.
We ensure that each instrument is swapped for the instrument labelled as the same SATB part in the target mixture, so that the target instruments play melodies that match their pitch range.

We sample $\set{\hat{z}_s}{N_s}$ from $\gauss{0}{\vrm{I}}$ conditioned on the manipulated set of timbre and pitch latents~(\neqref{eq:ldm_decoder}) and sample $\set{\hat{x}_s}{N_s}$ with $T=1000$ steps.
We expect that feeding $\vi{\hat{x}_s}$ to pre-trained instrument and pitch classifiers produces the targeted instrument and the original reference melody, respectively.
Because the mixture is obtained by $\hat{x}_m = \sum\nolimits_{i=1}^{N_s}\vi{\hat{x}_s}$ in the audio domain, the evaluation of disentanglement suggests the results of mixture rendering.
The results are reported in Table~\ref{tab:ldm_result}.

\begin{table}[t]
\small
\centering
\begin{tabular}{lrrrrrr}
\toprule
    & \multicolumn{2}{c}{Instrument ($\uparrow$)} & \multicolumn{2}{c}{Pitch ($\uparrow$)}     & \multicolumn{2}{c}{FAD ($\downarrow$)}       \\
\cmidrule(lr){2-3} \cmidrule(lr){4-5} \cmidrule(lr){6-7}
    & \multicolumn{1}{r}{Set} & \multicolumn{1}{r}{Single} & \multicolumn{1}{r}{Set} & \multicolumn{1}{r}{Single} & \multicolumn{1}{r}{Set} & \multicolumn{1}{r}{Single} \\
\midrule
M0  & 89.94          & \textbf{96.43}& 94.37          & \textbf{97.15} & 2.27          & \textbf{2.13} \\
M1  & \textbf{97.49} & 96.05          & 97.15          & 97.02          & \textbf{2.10} & 2.14          \\
M2   & 82.83          & 80.16          & \textbf{97.19}& 97.03          & 2.14          & 2.14          \\
\bottomrule
\end{tabular}
\caption{Disentanglement in terms of instrument and pitch classification (\%) and audio quality in terms of FAD.
}
\label{tab:ldm_result}
\end{table}

\subsubsection{Benchmark}
To justify the design that conditions the decoder with a set of source-level representations, we experiment with a model variant trained exclusively to reconstruct one source conditioned on a single source-level latent at a time.
Instead of rendering all sources in $T$ steps of diffusion sampling, this variant iteratively renders $N_s$ sources to recover a mixture, totalling $N_s \times T$ steps.
This could serve as a proxy to~\citet{lin_unified_2021} as they similarly use a query-based encoder to extract a pitch and a timbre latent of a single source and condition a decoder to reconstruct the source.

Moreover, we relax the training by discarding the $\vrm{BCE}$ loss in~\neqref{eq:dismix-loss} and evaluate whether the model can achieve decent performance without explicit pitch supervision.

\subsubsection{Disentanglement}
In Table~\ref{tab:ldm_result}, M0 is the single source-conditioned variant, M1 is the proposed set-conditioned model, and M2 drops the pitch supervision.
``Set'' and ``Single'' denote rendering $\set{\hat{x}_s}{N_s}$ conditioned on a set of source-level representations and a single condition for $N_s$ iterations, respectively.
The transformer accepts varying input sequence lengths, whereby all the models can operate with the two approaches regardless of how they were trained.

\begin{figure}[t]
    \centering
    \includegraphics[width=0.9\linewidth]{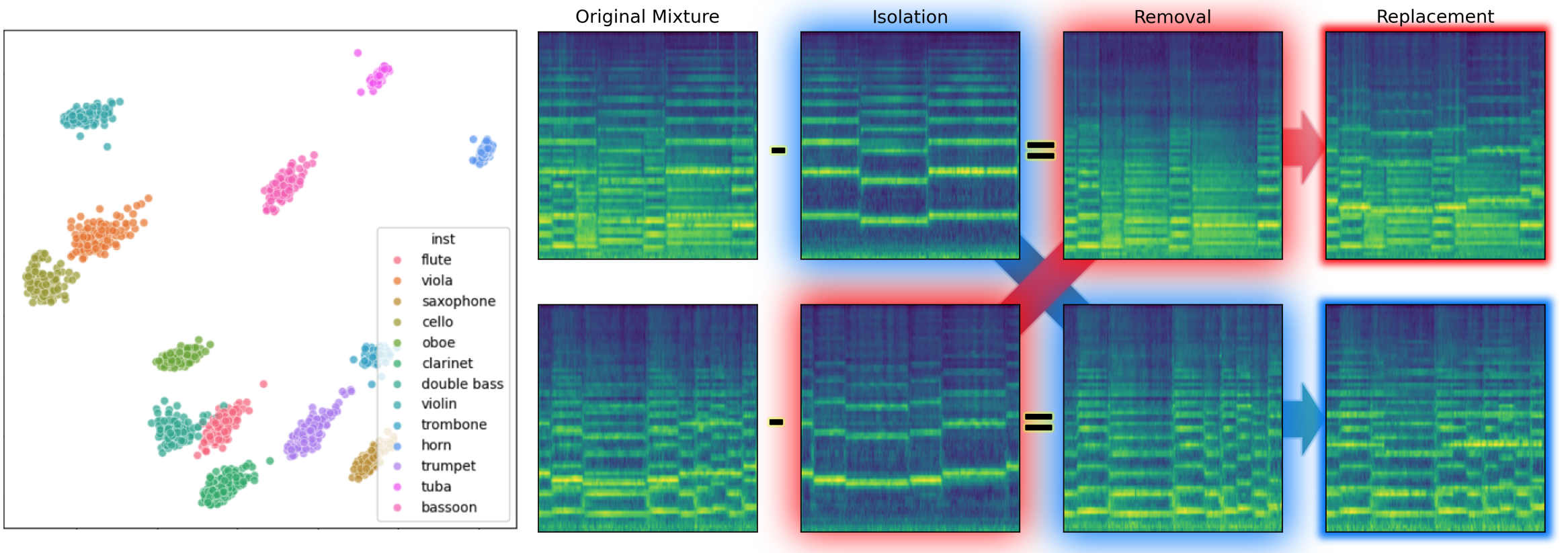}
    \caption{\textit{Left}: PCA of the timbre space. \textit{Right}: Compositional mixture rendering is achieved by modifying the members of a set of source-level latents.}
    \label{fig:teaser}
\end{figure}

M1 yields the best instrument classification accuracy at $97.49\%$ with the set-conditioned method, slightly outperforming M0 under the single source-conditioned setup at $96.43\%$.
M1's accurate instrument classification is reflected in the discriminative timbre space shown in Fig.~\ref{fig:teaser}.

Surprisingly, M0 yields decent performance using the set-conditioned approach ($89.94\%$ and $94.37\%$ for instrument and pitch accuracy) although it was not trained for the setup.

M2 achieves a decent performance without the pitch supervision, which implies the efficacy of the $\vrm{SB}$ layer proposed for disentanglement.
The inferior instrument accuracy suggests the pitch latent is contaminated by timbre.
Redundant information is preserved in the pitch latent for a good reconstruction, and thus M2 maintains a high pitch accuracy.

\subsubsection{Fr\'{e}chet Audio Distance (FAD)}
For each one of the $13$ instruments, we compute an FAD~\cite{kilgour2018fr} by comparing the VGG feature extracted from $\vi{\hat{x}_s}$ and that of the original data.
A successful timbre swapping and a good audio quality are necessary for achieving a low FAD.
Table~\ref{tab:ldm_result} reports the averaged FAD and suggests that the proposed M1 performs the best under the setup (Set) it was trained for.

Together, Table~\ref{tab:ldm_result} verifies the proposed design of the set-conditioned reconstruction by showing a better performance than its single source-conditioned counterpart.

\begin{figure}[t]
    \centering
    \includegraphics[width=0.9\linewidth]{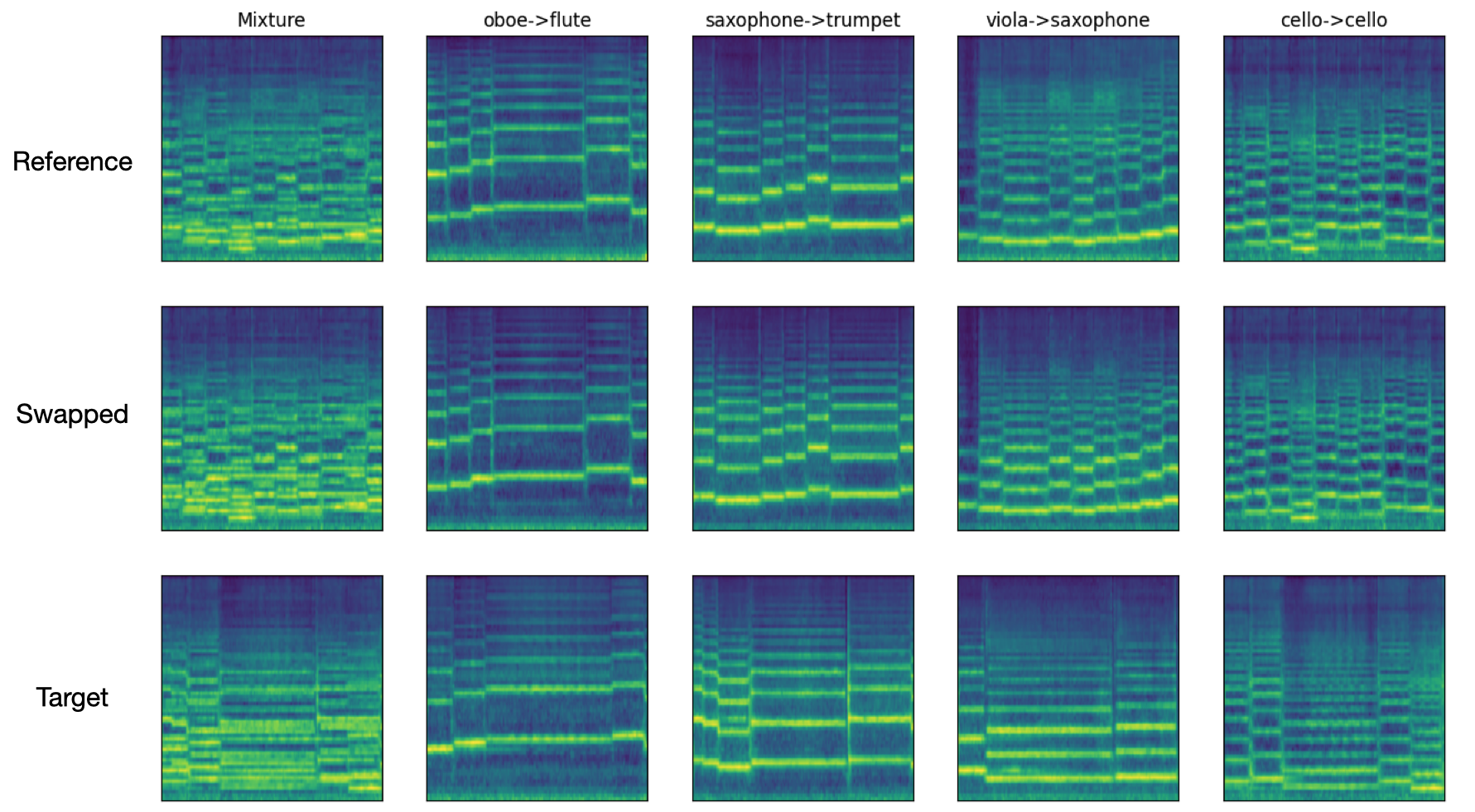}
    \caption{Replacing instruments of a reference mixture (the top row) given a target mixture (the bottom row).}
    \label{fig:swap}
\end{figure}

\subsubsection{Examples}
Fig.~\ref{fig:swap} shows an example of instrument replacement between a reference and a target mixture.
By replacing the timbre representations of the first three instruments for those from the target mixture, we can render a new mixture with the targeted instruments playing the original melodies.
We demonstrate audio samples at \texttt{yjlolo.github.io/dismix-audio-samples}.

\subsubsection{Compositional Mixture Rendering}
The right side of Fig.~\ref{fig:teaser} showcases an example of compositional mixture synthesis by modifying the members ($\vi{s}$) of a set of source-level representations, instead of manipulating the components ($\vi{\tau}$ and $\vi{\nu}$) of each member.

In the leftmost column, we begin with two-source mixtures for illustration.
We first extract the mixtures' source-level latents.
Then, by passing only one source to the decoder, we can isolate the source as shown in the second and the third column.
Finally, in the last column we can synthesise a new mixture conditioned on a set that takes a source from each of the two mixtures.
The two mixtures effectively ``exchange'' one of their sources.

%% file: sections/conclusion.tex
\section{Conclusion}
We present DisMix to tackle pitch-timbre disentanglement for multi-instrument mixtures.
The pitch and timbre latents act as modular building blocks, and we are able to render novel mixtures conditioned on sets of pitch and timbre representations.
In future work, we plan to relax the model by addressing the requirement for queries.
For applications, we consider ``attribute inpainting'' where the decoder is tasked to fill missing semantic information in its conditional inputs.


%

%% file: sections/appendix/appendix.tex
\newcommand{\prodi}{\prod_{i=1}^{N_s}}
\newcommand{\sumi}{\sum\nolimits_{i=1}^{N_s}}
\newcommand{\vx}[2]{#1^{(#2)}}

\section{Derivation of $\vcal{L}_{\vrm{ELBO}}$}\label{app:elbo}
For a multi-instrument mixture containing $N_s$ sources, we define its corresponding sets of the pitch annotations, the timbre latents, and the pitch latents as 
$\vcal{Y} = \set{y}{N_s}$, 
$\vcal{T} = \set{\tau}{N_s}$, and
$\vcal{V} = \set{\nu}{N_s}$,
respectively.
An instrument can be represented by a source-level representation $\vi{s} = \vrm{f}_{\theta_s}(\vi{\tau}, \vi{\nu})$, where $\vrm{f}_{\theta_s}$ is a deterministic function, the diamond node that precedes $x_m$ illustrated on the left side of Fig.~\ref{fig:plate}.
Similarly, the set of source-level representations is denoted as $\vcal{S} = \set{s}{N_s}$ and can be expressed in terms of $\vcal{T}$ and $\vcal{V}$ by $\vrm{f}_{\theta_s}(\vcal{T}, \vcal{V})$.

In the following derivation of the objective function $\vcal{L}_{\vrm{ELBO}}$ (\neqref{eq:elbo2}), we begin with the defined set notations, expand and simplify the equations with the 
factorised form of the posteriors~(\neqref{eq:posterior}), and recover $\vcal{L}_{\vrm{ELBO}}$ based on the deterministic posterior over the pitch latent~(\neqref{eq:pitch-encoder}).

\begin{align}
\log p(x_m  | \vcal{Y}) & \geq \vcal{L}_{\vrm{ELBO}} \label{eq:1}\\
={}&\E{
\qd{\tau}{\vcal{T} | x_m, \vcal{X}_q}
\qd{\nu}{\vcal{V} | x_m, \vcal{X}_q}
}{
\log \pd{m}{x_m | \vcal{S} = \vrm{f}_{\theta_s}(\vcal{T}, \vcal{V})}
} \label{eq:2}\\
& - \KLD{
\qd{\tau}{\vcal{T} | x_m, \vcal{X}_q} 
}{
p(\vcal{T})
}
- \KLD{
\qd{\nu}{\vcal{V} | x_m, \vcal{X}_q}
}{
\pd{\nu}{\vcal{V} | \vcal{Y}}
} \label{eq:3}\\ 
={}& \E{
\prodi\qd{\tau}{\vi{\tau} | x_m, \vi{x}_q}
\qd{\nu}{\vi{\nu} | x_m, \vi{x}_q}
}{
\log \pd{m}{x_m | \vcal{S} = \{\vrm{f}_{\theta_s}(\vi{\tau}, \vi{\nu})\}_{i=1}^{N_s}}
} \label{eq:4}\\
& - \int \prodi \qd{\tau}{\vi{\tau}| x_m, \vi{x}_q }
\log \frac{
\prodi \qd{\tau}{\vi{\tau}| x_m, \vi{x}_q }
}{
\prodi p(\vi{\tau})
} d\vx{\tau}{1}\ldots d\vx{\tau}{N_s} \label{eq:5}\\
& - \int \prodi \qd{\nu}{\vi{\nu}| x_m, \vi{x}_q }
\log \frac{
\prodi \qd{\nu}{\vi{\nu}| x_m, \vi{x}_q }
}{
\prodi \pd{\nu}{\vi{\nu} | \vi{y}}
} d\vx{\nu}{1}\ldots d\vx{\nu}{N_s}
\label{eq:6}\\
={} &  \E{
\prodi\qd{\tau}{\vi{\tau} | x_m, \vi{x}_q}
\qd{\nu}{\vi{\nu} | x_m, \vi{x}_q}
}{
\log \pd{m}{x_m | \vcal{S} = \{\vrm{f}_{\theta_s}(\vi{\tau}, \vi{\nu})\}_{i=1}^{N_s}}
} \label{eq:7}\\
& - \int \prodi \qd{\tau}{\vi{\tau}| x_m, \vi{x}_q }
 \sumi \log \frac{
\qd{\tau}{\vi{\tau}| x_m, \vi{x}_q }
}{
p(\vi{\tau})
} d\vx{\tau}{1}\ldots d\vx{\tau}{N_s}
\label{eq:8}\\
& - \int \prodi \qd{\nu}{\vi{\nu}| x_m, \vi{x}_q }
\sumi \log \frac{
\qd{\nu}{\vi{\nu}| x_m, \vi{x}_q }
}{
\pd{\nu}{\vi{\nu} | \vi{y}}
} d\vx{\nu}{1}\ldots d\vx{\nu}{N_s}
\label{eq:9}\\
={} & \E{
\prodi\qd{\tau}{\vi{\tau} | x_m, \vi{x}_q}
\qd{\nu}{\vi{\nu} | x_m, \vi{x}_q}
}{
\log \pd{m}{x_m | \vcal{S} = \{\vrm{f}_{\theta_s}(\vi{\tau}, \vi{\nu})\}_{i=1}^{N_s}}
} \label{eq:10}\\
& - \sumi \int \qd{\tau}{\vi{\tau}| x_m, \vi{x}_q }
 \log \frac{
\qd{\tau}{\vi{\tau}| x_m, \vi{x}_q }
}{
p(\vi{\tau}) 
} d\vi{\tau}
\label{eq:11}\\
& - \sumi \int \qd{\nu}{\vi{\nu}| x_m, \vi{x}_q }
\log \frac{
\qd{\nu}{\vi{\nu}| x_m, \vi{x}_q }
}{
\pd{\nu}{\vi{\nu} | \vi{y}}
} d\vi{\nu} 
\label{eq:12}\\
={} & \E{
\prodi\qd{\tau}{\vi{\tau} | x_m, \vi{x}_q}
\delta(\vi{\nu} - \vrm{f}_{\phi_\nu}(\vi{\hat{y}_{\vrm{bin}}}))
}{
\log \pd{m}{x_m | \vcal{S} = \{\vrm{f}_{\theta_s}(\vi{\tau}, \vi{\nu})\}_{i=1}^{N_s}}
} \label{eq:13}\\
& - \sumi\KLD{
\qd{\tau}{\vi{\tau}| x_m, \vi{x}_q }
}{
p(\vi{\tau})
} 
\label{eq:14}\\
& 
+ H(\delta(\vi{\nu} - \vrm{f}_{\phi_\nu}(\vi{\hat{y}_{\vrm{bin}}})))
+ \sumi
\int \delta(\vi{\nu} - \vrm{f}_{\phi_\nu}(\vi{\hat{y}_{\vrm{bin}}}))\log \pd{\nu}{\vi{\nu} | \vi{y}}
) d\vi{\nu} 
\label{eq:15}\\
={} & \E{
\prodi\qd{\tau}{\vi{\tau} | x_m, \vi{x}_q}
}{
\log \pd{m}{x_m | \vcal{S} = \{\vrm{f}_{\theta_s}(\vi{\tau}, \vi{\hat{\nu}})\}_{i=1}^{N_s}}
} \label{eq:16}\\
& - \sumi\KLD{
\qd{\tau}{\vi{\tau}| x_m, \vi{x}_q }
}{
p(\vi{\tau})
} 
+ \sumi
\log \pd{\nu}{\vi{\hat{\nu}} | \vi{y}}.
\end{align}
Starting with the standard form of the evidence lower bound (ELBO), we first assume that the posteriors over $\vcal{T}$ and $\vcal{V}$ are factorised in term~\eqref{eq:2}, and thus the two terms of the Kullback–Leibler divergence (KLD) corresponding to $\vcal{T}$ and $\vcal{V}$ are written as in term~\eqref{eq:3}.
In terms~\eqref{eq:4} to~\eqref{eq:6}, we further assume and factorise each posterior into a product over $N_s$ sources.
Note that the two KLD terms are expanded by the definition of KLD.
Based on the configuration of a deterministic posterior over $\vi{\nu}$ in \neqref{eq:pitch-encoder}, we rewrite $\qd{\nu}{\vi{\nu} | x_m, \vi{x}_q}$ as the Dirac delta function in terms~\eqref{eq:13} and~\eqref{eq:15}.
Finally, by replacing $\vi{\nu}$ with $\vi{\hat{\nu}} = \mathrm{f}_{\phi_\nu}(\vi{\hat{y}_{\vrm{bin}}}$), we recover $\vcal{L}_{\vrm{ELBO}}$ in~\neqref{eq:elbo2}.

\section{Implementation Details}
\subsection{The Simple Case Study}\label{app:ae}
\input{sections/appendix/ae_arch}

The mixture encoder $\vrm{E}_{\phi_m}$ and the query encoder $\vrm{E}_{\phi_q}$, mentioned in Section~\ref{sec:mixture_and_query}, 
share an architecture which is a stack of $\vrm{Conv1D}$ layers that take as input the mel spectrograms of $x_m$ and $\vi{x_q}$ and output $e_m, \vi{e_q}\in \R{64}$, respectively.
Table~\ref{tab:ae_conv1d} outlines the architecture.
The encoders are followed by a mean pooling along the temporal dimension such that a mel spectrogram of $\R{128 \times 10}$ is projected to $e_m, \vi{e_q}  \in \R{64}$.
The two are concatenated as a 128-dimensional vector used to extract the pitch and timbre according to \neqref{eq:posterior}.

By \neqref{eq:pitch-encoder} and given the concatenation of $e_m$ and $\vi{e_q}$, the pitch encoder $q_{\phi_\nu}$ first transcribes $\vi{\hat{y}} \in \R{N_p}$ using $\vrm{E}_{\phi_\nu}: \R{128} \rightarrow \R{N_p}$, where $N_p = 52$.
$\vi{\nu} \in \R{64}$ is then extracted by applying $\vrm{f}_{\phi_\nu}: \R{N_p} \rightarrow \R{64}$ to $\vi{\hat{y}_{\vrm{bin}}}$.
Table~\ref{tab:ae_pitch} shows the complete architecture.
The timbre encoder shares the same architecture as $\vrm{E}_{\phi_\nu}$, except for that the last layer is replaced by a Gaussian parameterisation layer which consists of two linear layers that project the $256$-dimensional hidden state to the mean and variance of the posterior, from which the timbre latent $\vi{\tau} \in \R{64}$ is sampled.

The $\vrm{FiLM}$ layer $\vrm{f}_{\theta_s}$ obtains the source-level latent $\vi{s}$ by combining the pitch and the timbre latent.
In particular, the timbre latent is linearly transformed to compute the scaling and shifting factors, and the pitch latent is modulated by the factors.

To reconstruct the mel spectrograms, the source-level latent $\vi{s}$ is then temporally broadcast to match the number of time frames $10$ of the mel spectrograms.
A two-layer bi-directrional gated recurrent unit (GRU) then transforms the broadcast $\vi{s}$ to an output of $\R{128 \times 10}$ which is then processed by a linear layer to reconstruct the input mel spectrograms.

\subsubsection{Pitch Priors}
The factorised pitch prior $\mu_{\theta_\nu}^{\vrm{fac}}$ takes as input the pitch annotation $\vi{y}$ and reuses $\vrm{f}_{\phi_\nu}$ for the transformation.
The rest of the network shares the same architecture as the timbre encoder as it also parameterises a Gaussian $\pd{\nu}{\vi{\nu} | \vi{y}} = \vcal{N}(\vi{\nu}; \mu_{\theta_\nu}^{\vrm{fac}}(\vi{y}), \sigma_\nu^2 \vrm{I})$, except for that the input size is $64$.
The standard deviation $\sigma_\nu$ is fixed at $0.5$.

\input{sections/appendix/transformer_diagram}

For the expressive variant, each element from the set of pitch annotations $\vcal{Y}_{\backslash{i}}$ is also transformed by $\vrm{f}_{\phi_\nu}$ first.
To handle a set of inputs, we implement a transformer.
The architecture consists of three blocks of a regular post-norm transformer, as shown in Fig.~\ref{fig:post-norm-transformer}.
We use a four-head attention and an embedding size of $64$.
The feed-forward block is a two-layer $\vrm{MLP}$ with a $\vrm{ReLU}$ and a size of $64$.

The elements in the set $\vcal{Y}_{\backslash{i}}$ are treated as a sequence of tokens and fed to the transformer without adding positional encodings to preserve permutation invariance.
A max pooling is applied to collapse the $N_s -1$ tokens followed by a Gaussian parameterisation to match the dimension of $\vi{\nu}$.
The Gaussian layer consists of $2 \times K$ linear layers to parameterise the mean and standard deviation of $K$ components in a Gaussian mixture.

\subsubsection{Latent Linearity}\label{app:ae_lin}
We introduce~\neqref{eq:mixture_prior} to avoid implementing a potentially complicated decoder, and thereby admit a  linearity between the sources and the mixture, represented by $s_{\vrm{sum}} = \sumi\vi{s}$ and $e_m$, respectively.
This linearity could allow for manipulating specific sources without extracting the entire set of source-level representations as follows, which we leave for future work.
Because the decoder $\theta_m$ would accept $e_m$ in addition to $\set{s}{N_s}$ for reconstructing $x_m$, we can manipulate a specific source $i$ and sample a novel mixture by first computing a residual $r = e_m - \vi{s}$, manipulating $\vi{s}$ to get $\vi{\hat{s}}$, and finally using $\hat{e_m} = r + \vi{\hat{s}}$ to render a novel mixture.

\subsubsection{Evaluation}
The instrument and the pitch classifier share the same architecture, except for their last layers whose sizes are the numbers of instruments $3$ and the pitches $52$, respectively.
Both start with the architecture outlined in Table~\ref{tab:ae_conv1d} followed by a three-layer $\vrm{MLP}$.
The input and output sizes of the first two layers are $64$ with the $\vrm{ReLU}$ activation, while that of the last layer are $64$ and the numbers of classes described above.

\input{sections/appendix/ldm_arch}
\subsection{The Latent Diffusion Model}\label{app:ldm}
\subsubsection{Encoders}

We extract mel spectrograms of $\R{64 \times 400}$ according to the parameters specified in Section~\ref{sec:coco}, where the dimensions correspond to the frequency and time axes, respectively.
The mixture encoder $\vrm{E}_{\phi_m}$ processes the mel spectrograms of $x_m$ and outputs $e_m \in \R{8 \times 100 \times 16}$, with the dimensions corresponding to the channels, time frames, and feature size, respectively.
We use and freeze $\vrm{E}_{\vrm{vae}}$, the pre-trained encoder of the VAE in AudioLDM2~\cite{audioldm2-2024taslp} as the mixture encoder.
Similarly, we reuse the architecture of $\vrm{E}_{\vrm{vae}}$ for the query encoder $\vrm{E}_{\phi_q}$ but train it from scratch and append a temporal pooling layer, which outputs $\vi{e_q} \in \R{8 \times 1 \times 16}$.
To combine the information of $\vi{e_q}$ and $e_m$ for extracting the pitch and timbre latents according to~\neqref{eq:posterior}, $\vi{e_q}$ is broadcast and concatenated with $e_m$ along the feature dimension.
The outcome is then transformed back to the original feature dimension of $16$ by a $1 \times 1$ $\vrm{Conv2D}$ filter before being passed for pitch and timbre extraction.

The pitch encoder first transcribes with $\vrm{E}_{\phi_\nu}: \R{8 \times 100 \times 16} \rightarrow \R{129 \times 400}$ whose architecture is based on the decoder $\vrm{D}_{\vrm{vae}}$ from the pre-trained VAE in AudioLDM2.
We modify the output dimension from $64$ to $129$ to accommodate the number of pitch values in the CocoChorale dataset~\cite{wu2022chamber} and train $\vrm{E}_{\phi_\nu}$ from scratch.
The output of $\vrm{E}_{\phi_\nu}$ is then binarised by the $\vrm{SB}$ layer and passed through  $\vrm{f}_{\vrm{\phi_\nu}}: \R{129 \times 400} \rightarrow \R{8 \times 100 \times 16}$ to derive the pitch latent $\vi{\nu}$, where $\vrm{f}_{\vrm{\phi_\nu}}$ is based on the architecture of $\vrm{E}_{\vrm{vae}}$ and is trained from scratch.

The timbre encoder $q_{\phi_q}: \R{8 \times 100 \times 16} \rightarrow \R{8 \times 1 \times 16}$ converts the input which concatenates $e_m$ and $\vi{e_q}$ and outputs the timbre latent $\vi{\tau}$, a time-invariant representation that matches the dimension of $\vi{e_q}$.
Its architecture is specified in Table~\ref{tab:ldm_timbre_encoder} which is followed by a temporal pooling layer.

\subsubsection{Partition}
We further elaborate the partition method described in Section~\ref{sec:instance_2}.
Recall that the diffusion target $z_{m, 0}$ in~\neqref{eq:ldm_decoder} is defined in terms of $z_{s, 0} \in \R{100 \times (16 \times 8)}$ extracted from $\vrm{E}_{\vrm{vae}}$, where we have transposed and flattened the extracted feature which had the dimensions $8 \times 100 \times 16$, corresponding to channels, time frames, and feature size, respectively.
We partition $z_{s,0}$ along the time axis into $L=25$ patches and flatten each patch by $\vrm{Par}(\vi{z_s}): \R{100 \times (16 \times 8)} \rightarrow \R{25 \times D_z'}$, where $D_z' = \frac{100}{25} \times 16 \times 8 = 512$.
By repeating the same process for all $N_s$ sources, we obtain and define $z_{m, 0} \in \R{(N_s \times 25) \times D_z'}$.
We can understand it as a sequence of $N_s \times 25$ patches with the size of each patch being $D_z' = 512$.

The timbre latent is broadcast along the time axis and concatenated with the pitch latent to form a source-level latent $\vi{s} \in \R{8 \times 100 \times 32}$. 
We follow the same process above, repeated for $N_s$ sources, to obtain $s_c \in \R{(N_s \times 25) \times D_s'}$, where $D_s' = \frac{100}{25} \times 32 \times 8 = 1024$.
Similarly, it is a sequence of $N_s \times 25$ patches with the size of each patch being $D_s' = 1024$.

\subsubsection{Decoder}
The transformer $p_{\theta_m}$ we implement for the LDM in~\neqref{eq:ldm_decoder} consists of three regular transformer blocks illustrated in Fig.~\ref{fig:post-norm-transformer} and we set the number of heads to four for the multi-head self-attention mechanism.
The embedding size is the same as the input size $D_z'$.
As described in Section~\ref{sec:instance_2}, we replace the layer normalisation layer with its adaptive counterpart to condition the input $z_{m, t}$ with $s_c$.

\subsubsection{Evaluation}
Table~\ref{tab:ldm_inst_clfr} outlines the architecture of the instrument classifier used for the evaluation of disentanglement.
The last layer is linearly projected to the logit of the $13$ classes of the instruments.
For the pitch classification, we employ Crepe, the state-of-the-art monophonic pitch extractor~\cite{kim2018crepe}.\footnote{\texttt{https://github.com/maxrmorrison/torchcrepe}}
We use a public repository~\footnote{\texttt{https://github.com/gudgud96/frechet-audio-distance/tree/main}} to measure the FAD.

%% file: sections/appendix/ae_arch.tex
\begin{table}[ht!]
\centering
\begin{tabular}{@{}ccccccc@{}}
\toprule
\textbf{Input channel} & \textbf{Output channel} & \textbf{Kernel size} & \textbf{Stride} & \textbf{Padding} & \textbf{Normalization} & \textbf{Activation} \\ \midrule
128                    & 768                     & 3                    & 1               & 0                & Layer                  & ReLU                \\
768                    & 768                     & 3                    & 1               & 1                & Layer                  & ReLU                \\
768                    & 768                     & 4                    & 2               & 1                & Layer                  & ReLU                \\
768                    & 768                     & 3                    & 1               & 1                & Layer                  & ReLU                \\
768                    & 768                     & 3                    & 1               & 1                & Layer                  & ReLU                \\
768                    & 64                      & 1                    & 1               & 1                & None                   & None                \\ \bottomrule
\end{tabular}
\caption{The $\vrm{Conv1D}$ layers used to implement $\vrm{E}_{\phi_m}$ and $\vrm{E}_{\phi_q}$ for the simple model in Section~\ref{sec:instance_1}.}
\label{tab:ae_conv1d}
\end{table}

\begin{table}[ht!]
\centering
\begin{tabular}{@{}llccc@{}}
\toprule
\textbf{Module} & \textbf{Input size} & \textbf{Output size} & \textbf{Normalization} & \textbf{Activation} \\ \midrule
\multirow{7}{*}{$\mathrm{E}_{\phi_\nu}$} & 128 & 256  & Layer & ReLU    \\
 & 256 & 256  & Layer & ReLU    \\
 & 256 & 256  & Layer & ReLU    \\
 & 256 & 256  & Layer & ReLU    \\
 & 256 & 256  & Layer & ReLU    \\
 & 256 & 256  & Layer & ReLU    \\
 & 256 & 52   & None  & Sigmoid \\ \midrule
\multicolumn{5}{c}{Stochastic Binarisation (SB)} \\ \midrule
\multirow{3}{*}{$\mathrm{f}_{\phi_\nu}$} & 52  & 64   & None  & ReLU    \\
 & 64  & 64   & None  & ReLU    \\
 & 64  & 64   & None  & None    \\ \bottomrule
\end{tabular}
\caption{
The three modules of the pitch encoder for the simple model in Section~\ref{sec:instance_1}: The transcriber $\vrm{E}_{\phi_\nu}$, the $\vrm{SB}$ layer, and the projector $\vrm{f}_{\phi_\nu}$.
}
\label{tab:ae_pitch}
\end{table}

%% file: sections/appendix/transformer_diagram.tex
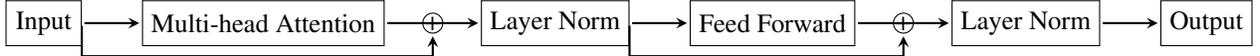
\begin{figure}[t!]
    \centering
    \begin{tikzpicture}[
        block/.style={rectangle, draw, minimum height=0.6cm, minimum width=1.0cm, align=center},
        arrow/.style={thick,->,>=stealth, shorten >=1pt, shorten <=1pt},
        sum/.style={circle, draw, inner sep=0pt, minimum size=0.6mm},
        layer/.style={draw, minimum height=0.6cm, minimum width=1.5cm, align=center}
    ]

    \node (input) [block] {Input};

    \node (mha) [layer, right=0.8cm of input] {Multi-head Attention};
    \draw [arrow] (input.east) -- (mha.west);

    \node (add1) [sum, right=0.5cm of mha] {+};
    \node (norm1) [block, right=0.5cm of add1] {Layer Norm};
    \draw [arrow] (mha.east) -- (add1.west) -- (norm1.west);
    \draw [arrow] (input.east) |- ([yshift=-0.3cm]add1.south) -- (add1.south);

    \node (ffn) [layer, right=0.8cm of norm1] {Feed Forward};
    \draw [arrow] (norm1.east) -- (ffn.west);

    \node (add2) [sum, right=0.5cm of ffn] {+};
    \node (norm2) [block, right=0.5cm of add2] {Layer Norm};
    \draw [arrow] (ffn.east) -- (add2.west) -- (norm2.west);
    \draw [arrow] (norm1.east) |- ([yshift=-0.3cm]add2.south) -- (add2.south);

    \node (output) [block, right=0.8cm of norm2] {Output};
    \draw [arrow] (norm2.east) -- (output.west);

    \end{tikzpicture}
    \caption{A regular post-norm Transformer block.}
    \label{fig:post-norm-transformer}
\end{figure}

%% file: sections/appendix/ldm_arch.tex
\begin{table}[t!]
\centering
\begin{tabular}{@{}ccccccc@{}}
\toprule
\textbf{Input channel} & \textbf{Output channel} & \textbf{Kernel size} & \textbf{Stride} & \textbf{Padding} & \textbf{Normalization} & \textbf{Activation} \\ \midrule
128                    & 128                     & 5                    & 2              & 0                & Group(1) & ReLU                \\
128                    & 128                     & 5                    & 2               & 0                & Group(1) & ReLU                \\
128                    & 128                     & 5                    & 2               & 0                & Group(1) & ReLU                \\
128                    & 256 & 1                    &  1              & 0                & None & None \\
\bottomrule
\end{tabular}
\caption{The $\vrm{Conv1D}$ layers used to implement the timbre encoder $q_{\phi_\tau}$ for the LDM in Section~\ref{sec:instance_2}.
The last row refers to the Gaussian layer where the $256$-dimensional output is split to represent the mean and standard deviation of the posterior.
The number in the parenthesis indicates the number of groups divided for normalisation.
}
\label{tab:ldm_timbre_encoder}
\end{table}

\begin{table}[b!]
\centering
\begin{tabular}{@{}ccccccc@{}}
\toprule
\textbf{Input channel} & \textbf{Output channel} & \textbf{Kernel size} & \textbf{Stride} & \textbf{Padding} & \textbf{Normalization} & \textbf{Activation} \\ \midrule
64                    & 64                     & 5                    & 2              & 0                & Group(1) & ReLU                \\
64                    & 64                     & 5                    & 2               & 0                & Group(1) & ReLU                \\
64                    & 64                     & 5                    & 2               & 0                & Group(1) & ReLU                \\
64                    & 16 & 1                    &  1              & 0                & None & None \\
\bottomrule
\end{tabular}
\caption{The instrument classifier used to evaluate the disentanglement of the proposed LDM.
The number in the parenthesis indicates the number of groups divided for normalisation.
}
\label{tab:ldm_inst_clfr}
\end{table}